\documentclass[prd,preprint,aps,superscriptaddress,preprintnumbers,letterpaper,natbib,floatfix,nofootinbib]{revtex4-1}

\usepackage{amssymb}
\usepackage{amsmath}
\usepackage{epsfig}
\usepackage{comment}
\usepackage{color}
\usepackage[hidelinks]{hyperref}
\usepackage{cleveref}
\usepackage{multirow}
\usepackage{capt-of}
\usepackage{siunitx}
\usepackage{graphicx}
\usepackage{slashed}
\usepackage{tabularx}
\usepackage{enumitem}
\usepackage{multirow}
\usepackage{booktabs}
\usepackage{isotope}
\usepackage{xcolor}
\usepackage{bm}
\usepackage{array,mathtools}
\usepackage{fontawesome}

\newcommand{\Neff}{N_\text{eff}}
\newcommand{\al}[1]{\begin{align} #1 \end{align} }

\newcommand{\gamp}{{\gamma '}}

\newcommand{\xxtopsps}{\chi \overline{\chi} \to \psi \overline{\psi}}

\newcommand\beq{\begin{eqnarray}}
\newcommand\eeq{\end{eqnarray}}

\usepackage{tikz,xcolor,hyperref}

\definecolor{lime}{HTML}{A6CE39}
\DeclareRobustCommand{\orcidicon}{\hspace{-1mm}
	\begin{tikzpicture}
		\draw[lime, fill=lime] (0,0) 
		circle [radius=0.16] 
		node[white] {{\fontfamily{qag}\selectfont \tiny \,ID}};
		\draw[white, fill=white] (-0.0525,0.095) 
		circle [radius=0.007];
	\end{tikzpicture}
	\hspace{-3mm}
}

\foreach \x in {A, ..., Z}{\expandafter\xdef\csname orcid\x\endcsname{\noexpand\href{https://orcid.org/\csname orcidauthor\x\endcsname}
		{\noexpand\orcidicon}}
}

\begin{document}

\title{Sub-MeV Dark Sink Dark Matter}

\author{Prudhvi N.~Bhattiprolu\orcidA}
\affiliation{Leinweber Center for Theoretical Physics, Department of Physics,\\ University of Michigan, Ann Arbor, MI 48109, USA}
\author{Robert McGehee\orcidB}
\affiliation{William I. Fine Theoretical Physics Institute, School of Physics and Astronomy,\\ University of Minnesota, Minneapolis, MN 55455, USA}
\author{Evan Petrosky\orcidC}
\affiliation{Leinweber Center for Theoretical Physics, Department of Physics,\\ University of Michigan, Ann Arbor, MI 48109, USA}
\author{Aaron Pierce\orcidD}
\affiliation{Leinweber Center for Theoretical Physics, Department of Physics,\\ University of Michigan, Ann Arbor, MI 48109, USA}

\begin{abstract}
 A Dark Sink uses dark-sector interactions to siphon energy from dark matter to lighter dark degrees of freedom, i.e. dark radiation. Here, we extend dark matter models containing a Dark Sink to sub-MeV masses. We consider a Dark Sink model where the dark matter is charged under a light dark photon that has kinetic mixing with the Standard Model. For sub-MeV dark matter masses, plasmon decays are the dominant mechanism for transferring energy to the dark sector. Relative to a standard freeze-in cosmology, reproducing the observed dark matter density in a Dark Sink structure requires an increase in the dark matter couplings to the Standard Model, and hence increased direct detection cross sections. These models provide benchmarks for current and upcoming direct detection experiments. Accounting for plasmon effects, we derive the range of possible dark matter masses and cross sections for Dark Sink models in the sub-MeV regime. We make code \href{https://github.com/prudhvibhattiprolu/FreezeIn}{\faGithub} available to reproduce our benchmarks; it may be of use for other freeze-in scenarios, including those where plasmon decays to the dark matter are important.
\end{abstract}

\maketitle
\preprint{FTPI-MINN-24-15}
\preprint{LCTP-24-13}

\tableofcontents
\newpage

\section{Introduction}
Particle dark matter with mass in the decades below a GeV has become a prime target of direct detection experiments.  For while ton-scale liquid noble gas experiments rapidly lose sensitivity at these masses, new technologies with lower energy thresholds hold the promise of extending to lower masses. But for dark matter with masses below an MeV, constraints on the energy density of relativistic degrees of freedom (parameterized by $\Neff$) are powerful~\cite{Planck:2018vyg,Sabti:2019mhn}. The strength of this constraint limits the number of dark matter candidates in this mass range that have both consistent cosmological histories and the promise of being directly detected in experiments. Some candidates assume highly non-trivial cosmological histories that involve, e.g., a late-time phase transition that alters the dark matter direct detection properties \cite{Elor:2021swj} or very low reheating temperatures \cite{Dror:2019onn,Dror:2019dib,Dror:2020czw,Bhattiprolu:2022sdd,Boddy:2024vgt}. Perhaps the simplest model of relatively light dark matter that may be directly detected couples to a light dark photon which kinetically mixes with the Standard Model (SM) photon and realizes its abundance via freeze-in~\cite{Dvorkin:2019zdi,Chang:2019xva}. 

And while the freeze-in benchmark is simple, it will take time for direct detection experiments to reach the relevant cross sections. In particular, though there are clever ways to probe the sub-MeV region at relatively small cross sections~\cite{An:2021qdl}, presently there are no experiments that probe the freeze-in benchmark. This must wait for future experiments based on novel target materials and detectors operating at exceptionally small energy thresholds (see e.g.~\cite{Hochberg:2015pha,Hochberg:2015fth,Hochberg:2017wce,Knapen:2017ekk,Knapen:2021bwg,Du:2022dxf}). So, understanding whether there are other minimal dark matter models with simple cosmologies that predict larger cross sections at such light masses is a timely question. As both current and future direct detection experiments push to probe ever lighter dark matter, it is important to know what consistent models may be discovered. 

A recent proposal has introduced benchmark models with cross sections ranging from the freeze-in values all the way up to current experimental bounds~\cite{Bhattiprolu:2023akk}. Such models could be discovered at any time.  In these models, the usual freeze-in benchmark is augmented by the presence of a ``Dark Sink."  This Dark Sink consists of new light fermions $\psi$ which behave as dark radiation and provide a dark matter annihilation channel, $\xxtopsps$. The $\psi$ thermalize with the dark matter $\chi$ and form a bath which is colder than the SM bath. The $\xxtopsps$ annihilation strength, independent of the strength of the dark matter-visible sector coupling, impacts the dark matter relic density. As this interaction strength increases, the required strength of interaction between dark matter and the visible sector must increase to compensate for dark matter lost to annihilation. The result is a larger scattering cross section off electrons, limited only by the requirement that the $\psi$'s do not contribute too much to $\Neff$. However, this constraint is not too restrictive \cite{Bhattiprolu:2023akk}.  Indeed, cross sections larger than those currently being probed at direct detection experiments~\cite{DarkSide:2022knj,PandaX:2022xqx,DAMIC-M:2023gxo,SENSEI:2023zdf} can be obtained. 

Initial work on the Dark Sink scenario \cite{Bhattiprolu:2023akk} focused on a dark matter mass range $\mathcal{O}(\text{MeV}) \lesssim m_\chi \lesssim \mathcal{O}(\text{TeV})$. In this paper we detail the consequences of adding a Dark Sink to the freeze-in benchmark for $m_\chi \lesssim \mathcal{O}(\text{MeV})$. This is qualitatively different due to the importance of plasmon decays in transferring energy from the SM bath to dark matter~\cite{Dvorkin:2019zdi}. We first review the Dark Sink model, then describe the calculation of the modified freeze-in accounting for both the presence of the Dark Sink  and the contribution of plasmon decays. We discuss the details of dark matter evolution, show the Dark-Sink interactions which yield the correct relic abundance, and find the range of possible cross sections relevant for current and future direct detection experiments.

\section{The Dark Sink Model}
As mentioned above, the Dark Sink model \cite{Bhattiprolu:2023akk} includes interactions between the dark matter $\chi$ and the visible sector as well as interactions within the dark sector. It is these latter interactions that allow energy transfer from $\chi$ (dark matter) to $\psi$ (dark radiation). For simplicity, we take the $\psi$ to be massless.

Interactions between the visible sector and the dark matter are provided via kinetic mixing between a new $U(1)^{\prime}$ gauge group and SM hypercharge. The dark matter $\chi$ is charged under this $U(1)^{\prime}$, while $\psi$ is neutral. We have
\begin{equation}
{\mathcal L} \supset -\frac{1}{4} \hat{X}_{\mu \nu} \hat{X}^{\mu \nu} + \frac{\epsilon_{Y}}{2} \hat{X}_{\mu \nu} \hat{B}^{\mu \nu},
\end{equation}
where $\hat{X}_{\mu \nu}$ is the field strength tensor for $U(1)^{\prime}$, $\hat{B}_{\mu \nu}$ is the field strength for SM hypercharge, and the hats on the fields emphasize that these are not mass eigenstates. After electroweak symmetry breaking, the kinetic mixing is
\begin{equation}
{\mathcal L} \supset \frac{\epsilon}{2} \hat{F}^{\mu \nu} \hat{X}_{\mu \nu},
\label{eq:portal}
\end{equation}
with $\epsilon = \epsilon_{Y} \cos \theta_{W}$ and $\theta_W$ the Weinberg angle. This kinetic mixing can be eliminated by a field redefinition of the SM photon. Under this redefinition, the SM electromagnetic current picks up small charges $\epsilon Qe$ under the dark photon, where $Qe$ is the charge of a given SM fermion. There is also a small kinetic mixing with the SM $Z$ boson, however, its effects are subdominant for the physics which we consider here, owing to the relatively small mass of the dark matter.

A particular combination of couplings
\begin{equation} \label{eq:kappadef}
\kappa \equiv \epsilon \sqrt{\frac{\alpha^{\prime}}{\alpha}},
\end{equation}
is relevant for both freeze-in and direct detection.  Here, $\alpha^{\prime}$ is the fine-structure constant of the $U(1)^{\prime}$, and $\alpha$ is the usual electromagnetic fine-structure constant.

Throughout this work, we take $\alpha^{\prime}$ sufficiently small so as to suppress dark matter self interactions. For a sufficiently light dark photon, the dark matter self-interaction cross section is in the classical regime and may be written as \cite{Tulin:2012wi}
\al{
\sigma_{\chi \chi} \approx \frac{16 \pi \alpha^{\prime 2}}{m_\chi^2 v^4} \log \left( \frac{m_\chi v^2}{2m_\gamp \alpha'} \right),
}
where $v$ is the dark matter relative velocity and $m_\gamp$ is the mass of the dark photon. For $10 \text{ keV} \lesssim m_\chi \lesssim 1 \text{ MeV}$, requiring that $\sigma_{\chi \chi}/m_\chi \lesssim 10 \text{ cm}^2/\text{g}$ \cite{Kaplinghat:2015aga} results in an upper bound on $\alpha'$ ranging from $10^{-13}$ to $10^{-10}$.

One might wonder whether the required smallness of $\alpha'$ places strong constraints on the allowed range of $\kappa$. However, for the above range of $\alpha'$ and $m_{\chi}$ for the largest values of $\kappa$ we consider below, the corresponding $\epsilon$ range from $10^{-3}$ to $10^{-5}$. This range of $\epsilon$ is unconstrained by, e.g., black hole superradiance~\cite{Baryakhtar:2017ngi,Siemonsen:2022ivj} or COBE/FIRAS~\cite{Fixsen:1996nj,Caputo:2020bdy} for a very light dark photon mass of $m_\gamp \sim 10^{-24} \text{ GeV}$. The dark photon also does not contribute to the dark matter density.

Within the dark sector, the interaction responsible for thermalizing the Dark Sink with dark matter is
\begin{equation}\label{eq:SinkInt}
{\mathcal L} \supset   \frac{y_{\chi} y_{\psi}}{m_{\Phi}^2} \overline{\chi} \chi \overline{\psi} \psi \equiv G_{\Phi} \overline{\chi} \chi \overline{\psi} \psi.
\end{equation}
Here, the $y_{i}$ represent Yukawa couplings of the dark fermions ($i = \chi, \psi$) to a new scalar $\Phi$ that is sufficiently heavy so that it is not produced on-shell during the early universe. The last equality defines the Dark Sink Fermi constant, $G_{\Phi}$.\footnote{Incidentally, we find this interaction does not induce a self-interaction cross section of a relevant size for current bounds.}

\section{Dark Matter Production}
The vector portal in Eq.~\eqref{eq:portal} allows dark matter pairs to be produced by both $2\rightarrow 2$ interactions, $\text{SM SM} \rightarrow \chi \overline{\chi}$, and plasmon decays, $\gamma^{\ast}\rightarrow \chi \overline{\chi}$. For the sub-MeV dark matter we focus on, the latter process dominates~\cite{Dvorkin:2019zdi,Chang:2019xva}. In this section, we first review dark matter production via plasmon decay, following \cite{Dvorkin:2019zdi}, taking advantage of the formalism of \cite{Braaten:1993jw}. We then turn to the set of coupled Boltzmann equations that i) describe the transport of energy density to the dark sector and ii) allow us to solve for the evolution of the dark matter number density. We first cover the contributions to freeze-in without the Dark Sink, and then discuss the relevant modifications in the presence of the Dark Sink. 

\subsection{Plasmon Properties}
\label{sec:plasmon}
\begin{figure}[t!]
    \includegraphics[width=0.8\columnwidth]{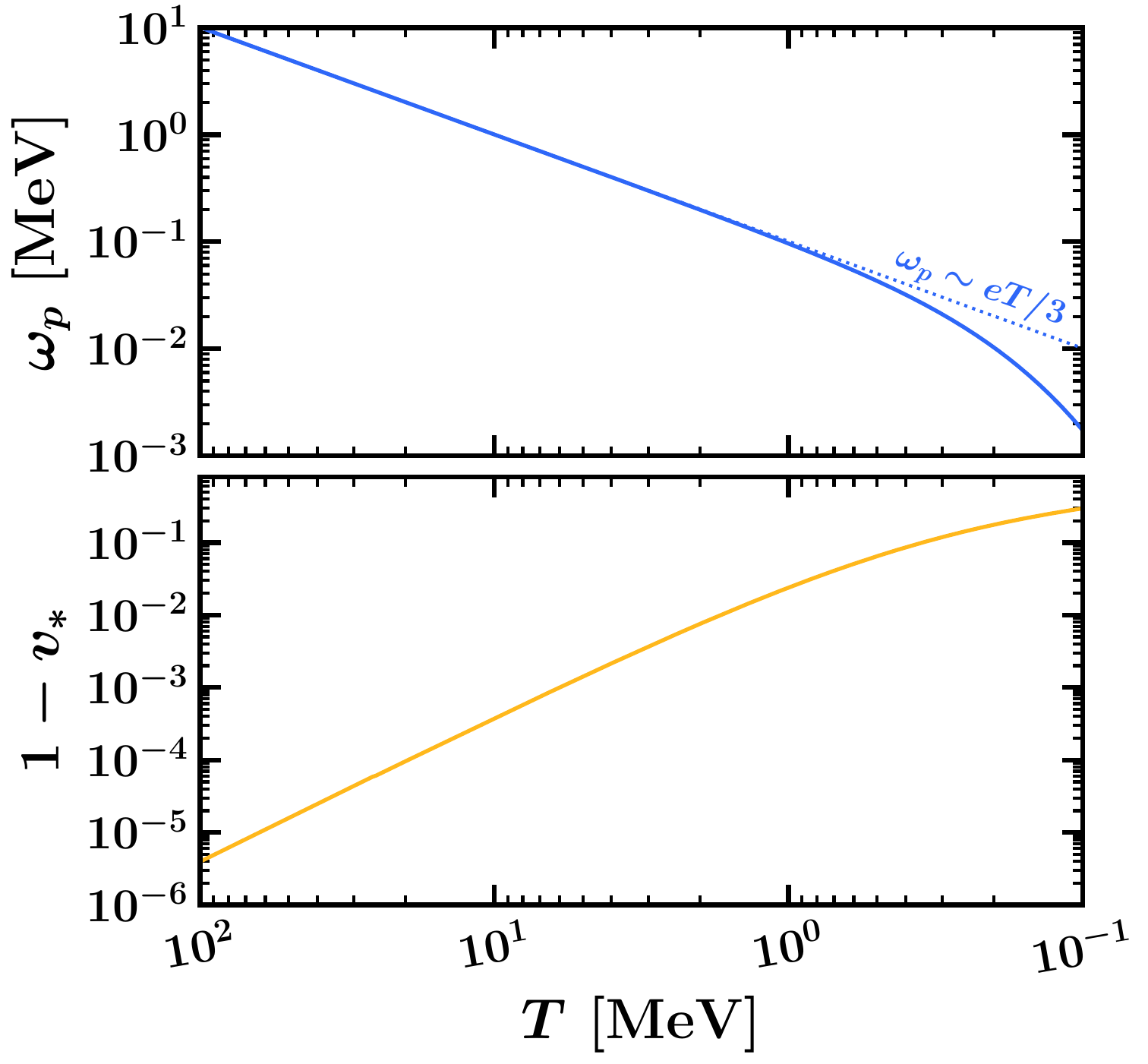}
 \caption{$\omega_p$ (top) and $1 - v_\ast$ (bottom) as functions of temperature. The plasma frequency, $\omega_p$, corresponds to the typical plasmon mass, and $v_\ast$ is the typical electron velocity in the plasma.}
\label{fig:PlasmonProperties}
\end{figure}

We begin by discussing the properties of the plasmons, following \cite{Braaten:1993jw}.  These properties will be important when we compute the decay rate of the plasmons to dark matter. We first define the plasma frequency $\omega_{p}$ that corresponds to the typical plasmon mass:
\beq
\omega_{p}^2 &=& \frac{2 e^2}{\pi^2}
\int_0^{\infty}{\frac{p^2 \, dp}{E} \left (1 - \frac{p^2}{3 E^2}\right) \frac{1}{1 + e^{E/T}}},
\eeq
where $E = \sqrt{p^2 + m_e^2}$. Another relevant scale is the typical velocity of electrons in the SM bath $v_{\ast} \equiv \omega_{1}/\omega_{p}$, where 
\beq
\omega_1^2 &=& \frac{2 e^2}{\pi^2}
\int_{0}^\infty
{\frac{ p^2 \, dp}{E}
 \left(\frac{5 p^2}{3E^2}-\frac{p^4}{E^4}\right) \frac{1}{1+ e^{E/T}}}
 .
\eeq
To get a feel for these quantities, we plot them in Fig.~\ref{fig:PlasmonProperties} as a function of temperature $T$. For temperatures much larger than the electron mass $m_{e}$, the plasma frequency is approximately given by 
\beq
    \omega_{p} &\approx& \frac{eT}{3}  \qquad (T \gg m_{e}).
\eeq
For temperatures well below $m_{e}$, the plasma frequency is exponentially suppressed. Throughout, we neglect the finite electron chemical potential which is only relevant for plasmon properties when $T \lesssim \mathcal{O}(20 \text{ keV})$ when $\omega_{p} \ll$ eV. This does not affect the dark matter abundance for any $m_{\chi}$ we consider.\footnote{It could however, matter for very light ($m_{\chi} \lesssim$ eV) sub-components of the dark matter~\cite{Iles:2024zka}.}

The masses of the transverse ($t$) and longitudinal ($\ell$) plasmons, denoted by $m_{i} \equiv \sqrt{\omega_{i}^2 - k^2}$ where $i= t, \ell$, may be calculated in terms of the above $\omega_p$ and $v_\ast$. Plasmon masses are obtained by solving
 \begin{eqnarray}
     \omega_{t}^2 &=& k^2 + \Pi_t ( \omega_{t}, k),\\
     \omega_{\ell}^2 &=& \frac{\omega_{\ell}^2}{k^2} \, \Pi_\ell ( \omega_{\ell}, k), \nonumber
 \end{eqnarray}
 with the polarization functions \cite{Silin:1960pya,Klimov:1982bv,Weldon:1982aq,Braaten:1993jw} 
 \begin{eqnarray}
 \Pi_{t} ( \omega, k)
 &=&
 \frac{3}{2} \omega_{p}^2\left[ \frac{\omega^2}{ v _{\ast}^2 k^2} - \frac{1}{2} \frac{\omega}{ v_{\ast} k} \left ( \frac{\omega^2}{ v _{\ast}^2 k^2}-1 \right) \ln{\frac{\omega + v_{\ast} k}{\omega - v_{\ast} k}}\right],\\
  \Pi_{\ell} ( \omega, k)
  &=&
  \frac{3 \omega_p^2}{v_\ast^2} \left(\frac{1}{2} \frac{\omega}{v_\ast k} \ln{\frac{\omega + v_{\ast} k}{\omega - v_{\ast} k}} - 1\right). \nonumber
 \end{eqnarray}

\begin{figure*}[t!]
  \begin{minipage}[]{0.8\columnwidth}
    \centering
    \includegraphics[width=\columnwidth]{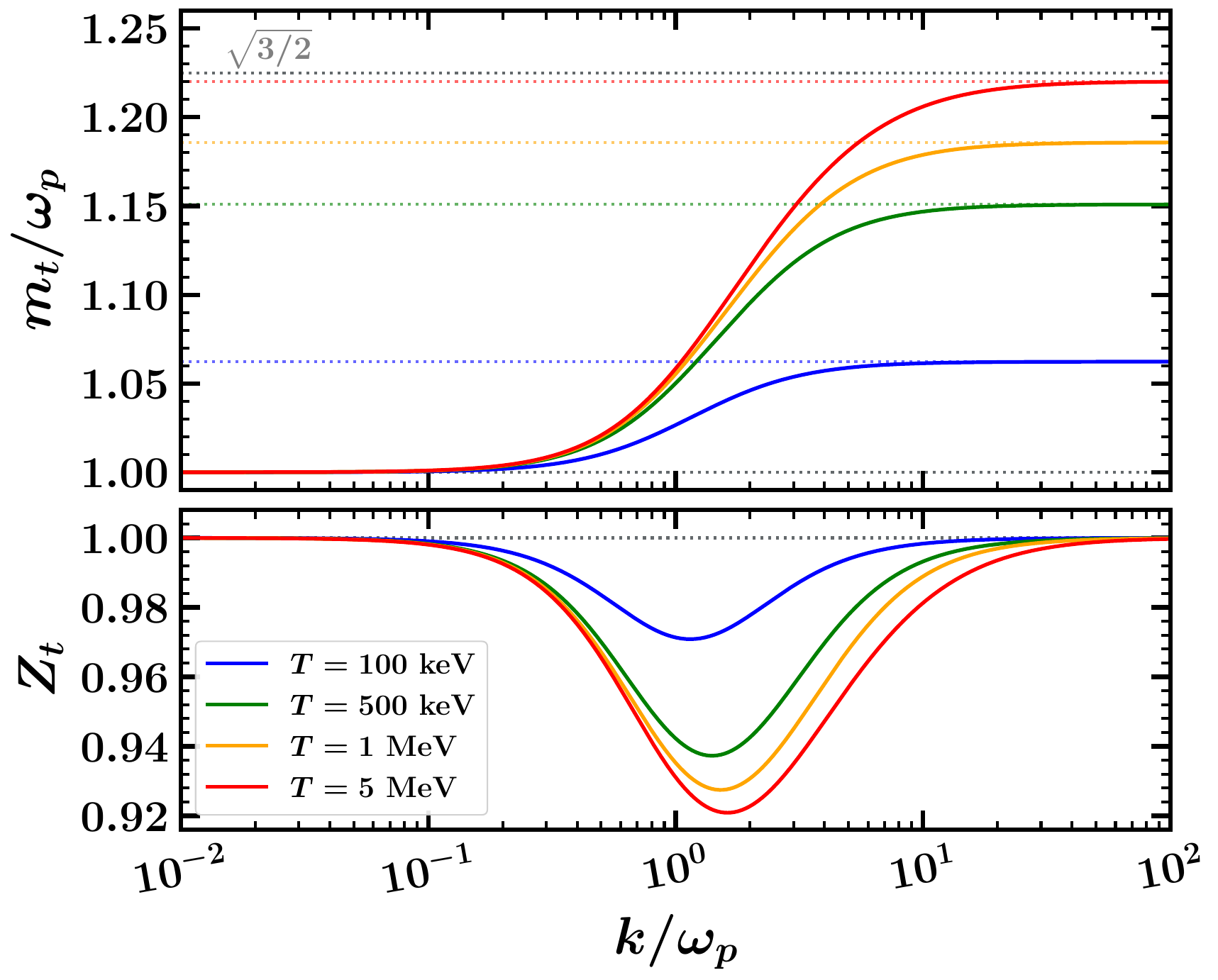}
  \end{minipage}
  \begin{minipage}[]{0.8\columnwidth}
    \centering
    \includegraphics[width=\columnwidth]{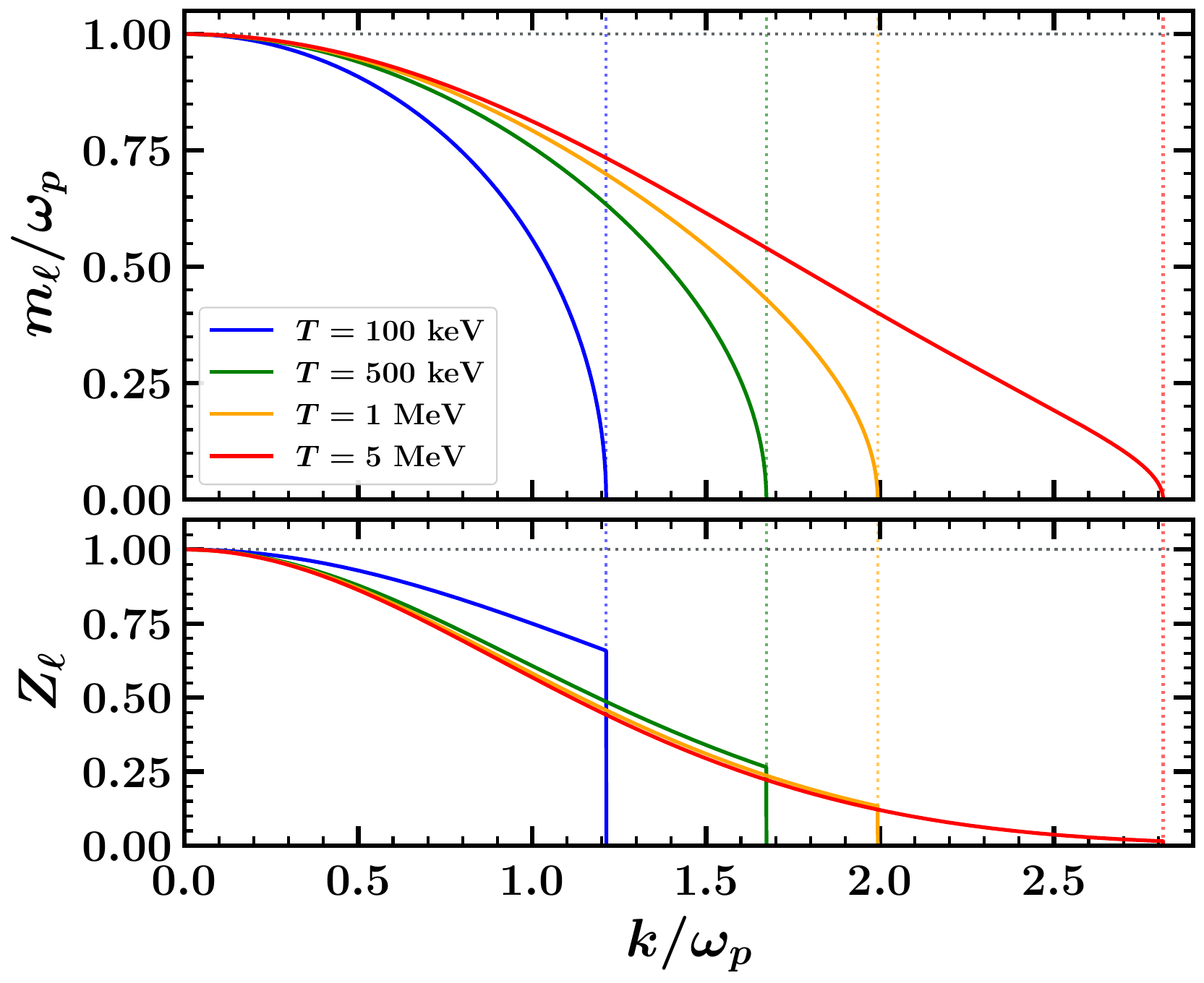}
  \end{minipage}
 \caption{The ratio of the plasmon mass $m_{t/\ell}$ to the plasmon frequency $\omega_p$ and the field strength renormalization factor $Z_{t/\ell}$ as functions of $k/\omega_p$, for various temperatures as labeled, for transverse ($t$) (top panel) and longitudinal ($\ell$) (bottom panel) plasmon modes.}
\label{fig:PlasmonMass}
\end{figure*}

The transverse plasmons propagate at all $k$, i.e., $0 \leq k < \infty$, while the longitudinal plasmons only propagate at low $k$, specifically, $0 \leq k < k_\text{max}$, with
\beq
k_\text{max} &=& \omega_p \, \sqrt{\frac{3}{v_\ast^2} \left(\frac{1}{2 v_\ast} \ln{\frac{1 + v_\ast}{1 - v_\ast}} - 1\right)}.
\eeq
Note that $k_\text{max} \rightarrow \infty$ in the relativistic limit where $v_\ast \rightarrow 1$ (or equivalently $m_e/T \rightarrow 0$). For the transverse mode, the plasmon mass increases as a function of $k$ from $\omega_p$ (at $k=0$) to $m_t^\text{max}$ (at $k = \infty$), given by
\beq
m_t^\text{max} = \omega_p \, \sqrt{\frac{3}{2 v_\ast^2} \left(1 - \frac{1 - v_\ast^2}{2 v_\ast} \ln{\frac{1 + v_\ast}{1 - v_\ast}}\right)},
\label{eq:mtmax}
\eeq
which asymptotes to $\sqrt{3/2} \, \omega_p$ in the relativistic limit. Conversely, for the longitudinal mode, the plasmon mass decreases as a function of $k$ from $\omega_p$ (at $k=0$) to $0$ (at $k=k_\text{max}$). The ratios of the plasmon masses $m_{\ell}$ and $m_{t}$ to $\omega_{p}$ as a function of $k/\omega_{p}$ are shown in Fig.~\ref{fig:PlasmonMass}. The cutoff momentum $k_\text{max}$ for the longitudinal modes is visible in the lower panels.

Just as the photon's mass gets renormalized in a plasma, so too does its wavefunction. These field strength renormalization factors $Z_{t/\ell}$ are:
\begin{eqnarray}
Z_{t} &=& \frac{2 \omega_{t}^2 ( \omega_{t}^2 - v_{\ast}^2 k^2)}{3 \omega_{p}^2 \omega_{t}^2 + (\omega_{t}^2 +k^2)(\omega_{t}^2 - v_{\ast}^2 k^2)- 2 \omega_{t}^2 ( \omega_{t}^2 -k^2)},\\
Z_{\ell} &=& \frac{2 (\omega_\ell^2 - v_\ast^2 k^2)}{3 \omega_p^2 - (\omega_\ell^2 - v_\ast^2 k^2)}. \nonumber
\end{eqnarray}
Numerical evaluation of these factors as a function of $k/\omega_{p}$ are also shown in Fig.~\ref{fig:PlasmonMass}. While $Z_{t}$ remains well within 10\% of unity for all values of $k/\omega_{p}$, $Z_{\ell}$ falls off rapidly as $k/\omega_{p}$ increases. Note that the results of Fig.~\ref{fig:PlasmonProperties} and Fig.~\ref{fig:PlasmonMass} are independent of the Dark Sink scenario.

An implementation of the plasmon properties discussed here, including masses and field strength renormalization factors, is made available with the {\sc FreezeIn} v2.0 package \cite{FreezeInCode}.

\subsection{Plasmon Decays}
\begin{figure}[t!]
    \includegraphics[width=0.8\linewidth]{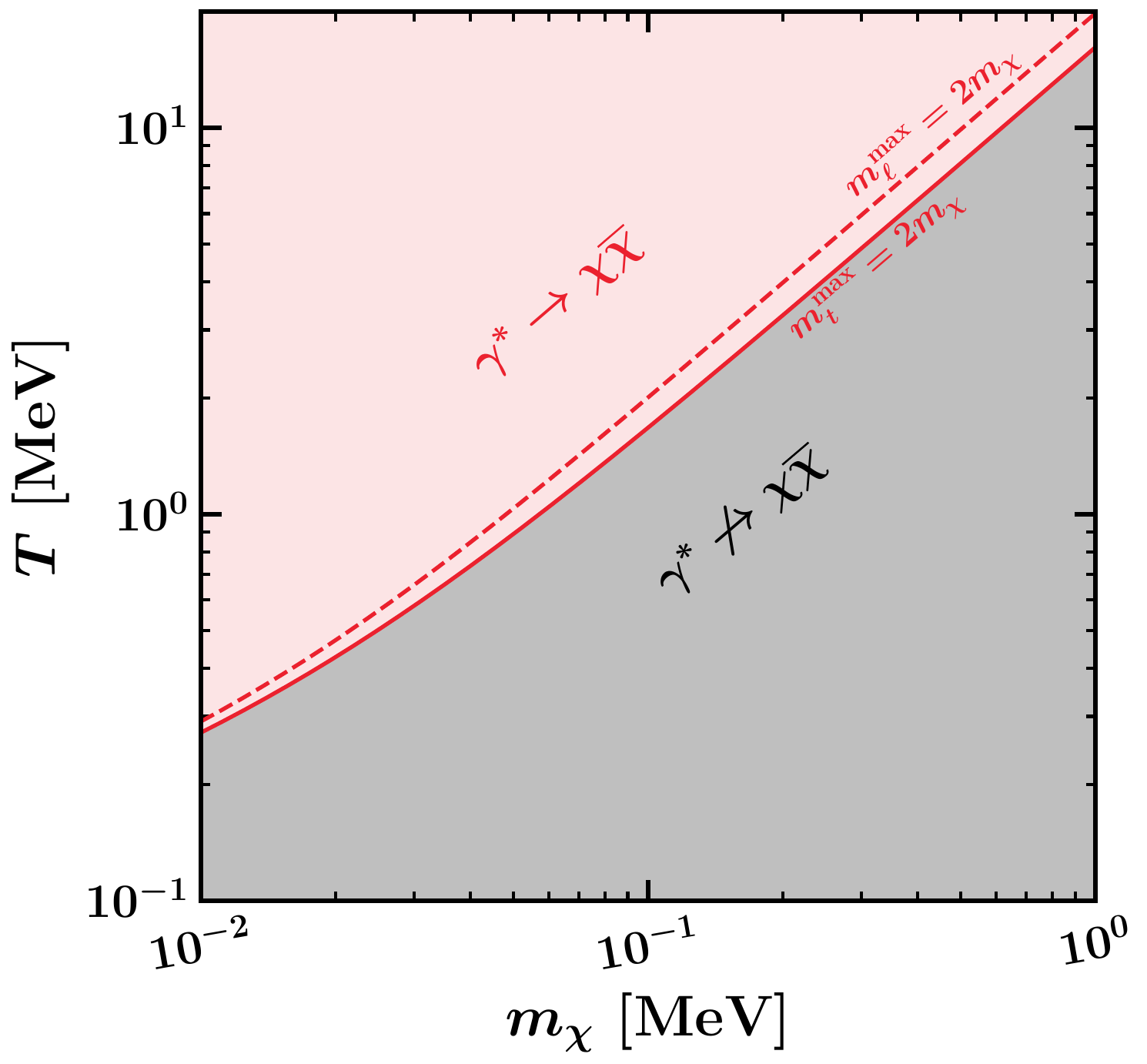}
     \caption{Temperatures where plasmon decays to dark matter are kinematically allowed as a function of the dark matter mass $m_{\chi}$. Whether decays are allowed is determined by the maximum value of the plasmon masses at a given $T$ (see Fig.~\ref{fig:PlasmonMass}). Solid and dashed red lines correspond to the temperatures at which the maximum value of the transverse plasmon mass, $m_t^\text{max}$ at $k=\infty$, and the longitudinal plasmon mass, $m_\ell^\text{max} \equiv \omega_p$ at $k=0$, equal $2 m_\chi$.}
\label{fig:PlasmonFITemperature}
\end{figure}

Armed with the expressions in the previous section, we can write down the thermally averaged plasmon decay rates as

\begin{eqnarray}
\label{eq:plasmondecay}
n_{\gamma_{t}^{\ast}} \langle \Gamma \rangle_{\gamma^{\ast}_{t}\rightarrow \chi \overline{\chi}}
&=&
\frac{\alpha \kappa^2}{ \pi^2} \int_{0}^{\infty}  \frac{k^2\, dk}{3} Z_{t} \frac{m_{t}^2}{\omega_{t} (e^{\omega_{t}/T}-1)}\left(1 + \frac{2 m_{\chi}^2}{m_{t}^2}\right) \sqrt{1 - \frac{4 m_{\chi}^2}{m_{t}^2}}
,\\
n_{\gamma_{\ell}^{\ast}} \langle \Gamma \rangle_{\gamma^{\ast}_{\ell}\rightarrow \chi \overline{\chi}}
&=&
\frac{\alpha \kappa^2}{2 \pi^2} \int_{0}^{k_\text{max}}  \frac{k^2\, dk}{3} Z_{\ell} \frac{\omega_{\ell}}{e^{\omega_{\ell}/T}-1}\left(1 + \frac{2 m_{\chi}^2}{m_{\ell}^2}\right) \sqrt{1 - \frac{4 m_{\chi}^2}{m_{\ell}^2}}. \nonumber
\end{eqnarray}
These expressions agree with the corrected published version of Ref.~\cite{Dvorkin:2019zdi}. In Fig.~\ref{fig:PlasmonFITemperature}, we show the temperatures where plasmons are kinematically allowed to decay to dark matter as a function of the dark matter mass $m_\chi$. These are the temperatures at which the maximum values of the plasmon masses, $m_t^\text{max}$ (at $k = \infty$) for transverse modes and $m_\ell^\text{max} \equiv \omega_p$ (at $k = 0$) for longitudinal modes, are greater than $2 m_\chi$.

In principle, decays of both longitudinal and transverse plasmons contribute to dark matter production. However, the contribution of the transverse plasmons dominates;  the decays of longitudinal plasmons make a sub-percent contribution over much of the parameter space of interest~\cite{Dvorkin:2019zdi}. Still, our numerical treatment includes both contributions.

\subsection{Boltzmann Equations without the Dark Sink}
The thermally averaged decay rates calculated above can be combined with the $2\rightarrow 2$ mechanism for dark matter production to write down a set of coupled Boltzmann equations. Since $\alpha'$ is so small, we may ignore $\chi \overline{\chi} \rightarrow \gamma^{\prime}\gamma^{\prime}$ interactions. The $2 \rightarrow 2$ process was discussed in detail in the context of the Dark Sink in Ref.~\cite{Bhattiprolu:2023akk}, but we reproduce the most important equations for reference. For the dark matter masses of interest ($<$ MeV), the most important $2 \rightarrow 2$ process is $e^{+} e^{-} \rightarrow \chi \overline{\chi}$. Neglecting the contribution from  $Z$-boson exchange--negligible for the masses of interest-- the $2 \rightarrow 2$ process has a fully-averaged squared matrix element given by\footnote{This notation differs slightly from that used in \cite{Bhattiprolu:2023akk}, where $\overline{|{\mathcal M}|}^2$ was defined to include the integral over $\cos \theta$ in the center-of-mass frame, and thus a factor of 2.}
\beq
\overline{|{\mathcal M}|}^2_{e^+ e^- \rightarrow \chi \overline{\chi}}
&=&
\frac{16}{3} \pi^2 \alpha^2 \kappa^2 \left(1 + \frac{2 m_e^2}{s}\right) \left(1 + \frac{2 m_{\chi}^2}{s}\right).
\eeq
This results in a thermally averaged cross section 
\beq
n_e^2 \langle \sigma v \rangle_{e^+ e^- \rightarrow \chi \overline{\chi}}
=
\frac{32 \,  T}{(4 \pi)^5}
\int_{s_\text{min}}^{\infty} ds \,
\overline{|{\mathcal M}|}^2_{e^+ e^- \rightarrow \chi \overline{\chi}}
\sqrt{1 - \frac{4 m_e^2}{s}} \sqrt{1 - \frac{4 m_{\chi}^2}{s}}
\sqrt{s}
K_1\left(\frac{\sqrt{s}}{T}\right),
\label{eq:therm2to2}
\eeq
where $s_\text{min} = \text{max}(4 m_e^2, 4 m_\chi^2)$, 
$n_e$ is the electron's equilibrium number density at temperature $T$, and $K_1$ is the modified Bessel function of the second kind with order 1. Since electrons are in equilibrium with the SM bath when they are relevant, we do not include an explicit ``eq'' label on their number density.

With this cross section and the thermally averaged plasmon decay rate of Eq.~(\ref{eq:plasmondecay}), we can write the Boltzmann equation for the dark matter yield $Y = n_{\chi}/s$ before the addition of the Dark Sink as \cite{Dvorkin:2019zdi}
\begin{eqnarray}
\label{eq:NoDarkSinkBoltzmann}
    -\overline{H}T s \frac{d Y}{d T} &=&
   n_e^2 \langle \sigma v \rangle_{e^+ e^- \rightarrow \chi \overline{\chi}} \left(1 - \frac{Y^2}{Y_{\mathrm{eq}}^2}\right) + n_{\gamma^{\ast}} \langle \Gamma \rangle_{\gamma^{\ast}\rightarrow \chi \overline{\chi}} \left(1 - \frac{Y^2}{Y_{\mathrm{eq}}^2}\right)\\
    &\approx& 
n_e^2 \langle \sigma v \rangle_{e^+ e^- \rightarrow \chi \overline{\chi}}  + n_{\gamma^{\ast}} \langle \Gamma \rangle_{\gamma^{\ast}\rightarrow \chi \overline{\chi}}, \nonumber
\end{eqnarray}
where $s$ is the entropy density and $Y_{\mathrm{eq}} = n_\chi^{\rm eq}/s$ is the equilibrium yield of $\chi$ at the SM bath temperature $T$. In the second line, we have made the approximation that the dark matter typically has an abundance well below the equilibrium one during freeze-in.  The quantity $\overline{H}$ is defined in terms of the Hubble expansion rate $H$ as 
\begin{equation}\label{eq:HbyHbarOrig}
    H/\overline{H} \equiv  1 + \frac{1}{3}\frac{d\ \mathrm{ln}g_{\ast,s}}{d\ \mathrm{ln}T}.
\end{equation}
For simplicity, we have only explicitly written the $2\to 2$ contribution coming from electron-positron annihilations, but we include all SM charged fermion annihilations in our numerical results. Note the total dark matter yield is $Y_{\chi+ \overline{\chi}}= 2 Y$. It is convenient to group together the plasmon and $2\rightarrow2$ contributions together by defining an effective thermally averaged cross section $ \langle \sigma v \rangle_{\mathrm{eff}}$ via 
\begin{equation}\label{eq:SigmaVEff}
    (n_\chi^{\rm eq})^2 \langle \sigma v \rangle_{\mathrm{eff}} \equiv n_e^2 \langle \sigma v \rangle_{e^+ e^- \rightarrow \chi \overline{\chi}} + n_{\gamma^{\ast}} \langle \Gamma \rangle_{\gamma^{\ast}\rightarrow \chi \overline{\chi}},
\end{equation}
where $n_\chi^{\rm eq}$ is the $\chi$'s equilibrium number density at SM bath temperature $T$. This quantity combines all processes involved in the creation of dark matter from SM particles. With this definition, the Boltzmann equation can be written 
\begin{equation}\label{eq:NoDarkSinkBoltzmann2}
\begin{aligned}
    -\overline{H}T s \frac{d Y}{d T} =
     (n_\chi^{\rm eq})^2\langle \sigma v \rangle_{\mathrm{eff}}\left(1 - \frac{Y^2}{Y_{\mathrm{eq}}^2}\right).
\end{aligned}
\end{equation}

\subsection{Adding the Dark Sink}
The addition of a Dark Sink allows for the thermalization of the dark sector and the depletion of the dark matter number density and energy density via the $\chi \overline{\chi} \rightarrow \psi \overline{\psi}$ interaction. A large enough $G_{\Phi}$ maintains the dark-sector temperature $T'$ during the epoch in which the relic density is produced. Quantities evaluated at the temperature of the dark-sector bath are noted by attaching a prime, e.g., $Y_{\mathrm{eq}}^{\prime} \equiv Y_{\mathrm{eq}} (T^{\prime})$. The entropy density of the Universe is comprised of both visible and dark-sector contributions:
\begin{equation}\label{eq:stotal}
    s = \frac{2\pi^2}{45}(g_{\ast,s}T^3 + g_{\ast,s}^{\prime}T^{\prime 3}) \equiv \frac{2\pi^2}{45}g_{\ast,s}^{\mathrm{eff}}T^3,
\end{equation}
where the last equality defines $g_{\ast,s}^{\mathrm{eff}}$.  To compute $T^{\prime}$, we solve the Boltzmann equation that governs energy transfer from the SM to the dark sector \cite{Bhattiprolu:2023akk} :
\beq \label{eq:rhoprimeBoltzmann}
-\overline{H} T \frac{d \rho^{\prime}}{dT} + 3 H (\rho^{\prime} + p^{\prime})
&=&
\sum_{i}
C^\rho_{i \rightarrow \chi \overline{\chi}} (T) -
C^\rho_{i \rightarrow \chi \overline{\chi}} (T^\prime)
\label{eq:energyDS}
\eeq
with
\beq
p^\prime = \frac{\rho^\prime}{3} - \frac{m_\chi^3 T^\prime}{3 \pi^2} K_1\left(\frac{m_\chi}{T^\prime}\right).
\eeq
Note the addition of the dark thermal bath changes the definition of $\overline{H}$ in Eq.~(\ref{eq:HbyHbarOrig}) to
\begin{equation}\label{eq:HbyHbarFull}
    H/\overline{H} \equiv  1 + \frac{1}{3}\frac{d\ \mathrm{ln}g_{\ast,s}^{\mathrm{eff}}}{d\ \mathrm{ln}T}+ \frac{1}{3}\frac{d\ \mathrm{ln}g_{\ast,s}^{\mathrm{eff}}}{d\ \mathrm{ln}T^{\prime}}\frac{T}{T^{\prime}}\frac{d T^{\prime}}{d T}
\end{equation}
where $g_{\ast,s}^{\mathrm{eff}}$ is defined above by Eq.~(\ref{eq:stotal}). We simplify the calculations and expressions here by assuming entropy conservation throughout, though some entropy creation occurs since heat flows from the SM bath to the dark bath. However, this is a tiny effect.  Since this is a freeze-in process, the rate of entropy transfer (and any related entropy non-conservation effect) to the dark sector is small. We have numerically verified including entropy non-conservation effects does not change any of our results by even $1\%$. 

The above equations are derived assuming Maxwell-Boltzmann statistics, and the sums are over all SM processes that can populate the dark sector i.e. $i = e^+ e^-, \gamma^\ast_t, \gamma^\ast_\ell, \ldots$.  The collision terms on the right-hand side are given by 
\beq
C^\rho_{e^+ e^- \rightarrow \chi \overline{\chi}} (T)
&=&
\frac{32 T}{(4 \pi)^5}
\int_{s_\text{min}}^\infty ds \,
\overline{|{\mathcal M}|}^2_{e^+e^- \rightarrow \chi \overline{\chi}}
\sqrt{s - 4 m_e^2} \sqrt{s - 4 m_\chi^2}
K_2\left(\frac{\sqrt{s}}{T}\right)
, \nonumber\\
C^\rho_{\gamma^\ast_t \rightarrow \chi \overline{\chi}} (T)
&=&
\frac{\alpha \kappa^2}{ \pi^2} \int_{0}^{\infty}  \frac{k^2\, dk}{3} Z_{t} \frac{m_{t}^2}{e^{\omega_{t}/T}-1}\left(1 + \frac{2 m_{\chi}^2}{m_{t}^2}\right) \sqrt{1 - \frac{4 m_{\chi}^2}{m_{t}^2}}
,\\
C^\rho_{\gamma^\ast_\ell \rightarrow \chi \overline{\chi}} (T)
&=&
\frac{\alpha \kappa^2}{2 \pi^2} \int_{0}^{\infty}  \frac{k^2\, dk}{3} Z_{\ell} \frac{\omega_{\ell}^2}{e^{\omega_{\ell}/T}-1}\left(1 + \frac{2 m_{\chi}^2}{m_{\ell}^2}\right) \sqrt{1 - \frac{4 m_{\chi}^2}{m_{\ell}^2}}. \nonumber
\label{eq:collisionPlasmon}
\eeq
In the above, $K_{2}$ is a modified Bessel function of the second kind with order 2. Solving Eq.~(\ref{eq:rhoprimeBoltzmann}) gives the dark-sector temperature $T^{\prime}$ as a function of the dark matter parameters and temperature of the SM bath, i.e. $T^{\prime} = T^{\prime}(T, \kappa, m_{\chi})$. 

Upon including the new dark-sector interaction, the dark matter Boltzmann equation becomes (cf. Eq.~(\ref{eq:NoDarkSinkBoltzmann2})), 
\begin{eqnarray}
\label{eq:FullBoltzmann}
    -\overline{H}T s \frac{d Y}{d T} &=&    
    (n_\chi^{\rm eq})^2
    \langle \sigma v \rangle_{\mathrm{eff}}\left(1 - \frac{Y^2}{Y_{\mathrm{eq}}^2}\right) + (n_\chi^{\mathrm{eq} \,\prime})^{2} \langle \sigma v \rangle_{\chi \overline{\chi} \to \psi \overline{\psi}}^{\prime}\left(1 - \frac{Y^2}{Y_{\mathrm{eq}}^{\prime 2}}\right).
\end{eqnarray}
Here, $n_\chi^{\rm eq\,\prime}$ is the $\chi$'s equilibrium number density at temperature $T'$. In the above equation, unprimed yields, number densities, and thermally averaged cross sections are evaluated at the SM bath temperature; primed yields, number densities, and thermally averaged cross sections are evaluated at the dark-sector bath temperature. The effect of the Dark Sink is to add a term to the right-hand side of the Boltzmann equation. The thermally averaged cross section for $\chi \overline{\chi} \to \psi \overline{\psi}$ is 
\beq
(n_\chi^{\mathrm{eq} \,\prime})^{2}\langle \sigma v \rangle_{\chi \overline{\chi} \to \psi \overline{\psi}}^{\prime}
&=&
\frac{G_\Phi^2}{2} \frac{16 T^\prime}{(4 \pi)^5} \int_{4 m_\chi^2}^\infty ds \, s^{5/2} \left(1 - \frac{4 m_\chi^2}{s}\right)^{3/2} K_1\left(\frac{\sqrt{s}}{T^\prime}\right).
\eeq
In the limit $m_\chi/T^\prime \gg 1$,  of interest for understanding the  evolution of the dark matter yield,
\beq
\langle \sigma v \rangle_{\chi \overline{\chi} \to \psi \overline{\psi}}^{\prime} &\approx& \frac{3}{4 \pi} \, G_\Phi^2 m_\chi T^\prime.
\label{eq:sigmavprimelargex}
\eeq

To compare the dark matter creation and depletion rates, it is convenient to define the quasi-static equilibrium yield $Y_{\mathrm{QSE}}$ (following the nomenclature of \cite{Chu:2011be})

\begin{equation}\label{eq:yieldQSE}
    Y_{\mathrm{QSE}} \equiv Y_{\mathrm{eq}} \, \sqrt{\frac{\langle \sigma v \rangle_{\mathrm{eff \; \; \; \; \; \; \; \; }}}{\langle \sigma v \rangle_{\chi \overline{\chi} \to \psi \overline{\psi}}^{\prime}}}.
\end{equation}
This quantity is small when the cross section for depletion of dark matter into dark radiation dominates that for the creation of dark matter from SM particles.  When these cross sections are equal, $Y_\text{QSE}$ is equal to the equilibrium yield at the temperature of the SM bath. In terms of $Y_\text{QSE}$, the full Dark Sink Boltzmann equation, Eq.~(\ref{eq:FullBoltzmann}) can be written 
\beq \label{eq:FullBoltSimp}
     -\frac{\overline{H}T}{s} \frac{d Y}{d T} &=& \langle \sigma v \rangle_{\chi \overline{\chi} \to \psi \overline{\psi}}^{\prime} \left[ Y_{\mathrm{eq}}^{\prime 2} - Y^2 + \left(1 - \frac{Y^2}{Y_{\mathrm{eq}}^{ 2}}\right)Y_{\mathrm{QSE}}^2\right].
\eeq
Below, we discuss the impact of adding a Dark Sink in detail and explore various special cases of the equation. 

\section{Results\label{sec:results}}

In this section, we first discuss the evolution of the dark matter density in more detail.  We then show results on the required size of the Dark Sink interactions required to achieve the observed dark matter density, paying attention to the impact of plasmon decays.  Finally, we present the range of possible direct detection cross sections in this Dark Sink model; these benchmarks are a main result of this work.

\subsection{Dark Matter Evolution}
\label{sec:evolution}

\begin{figure*}[t!]
  \begin{minipage}[]{\columnwidth}
    \centering
    \includegraphics[width=0.8\columnwidth]{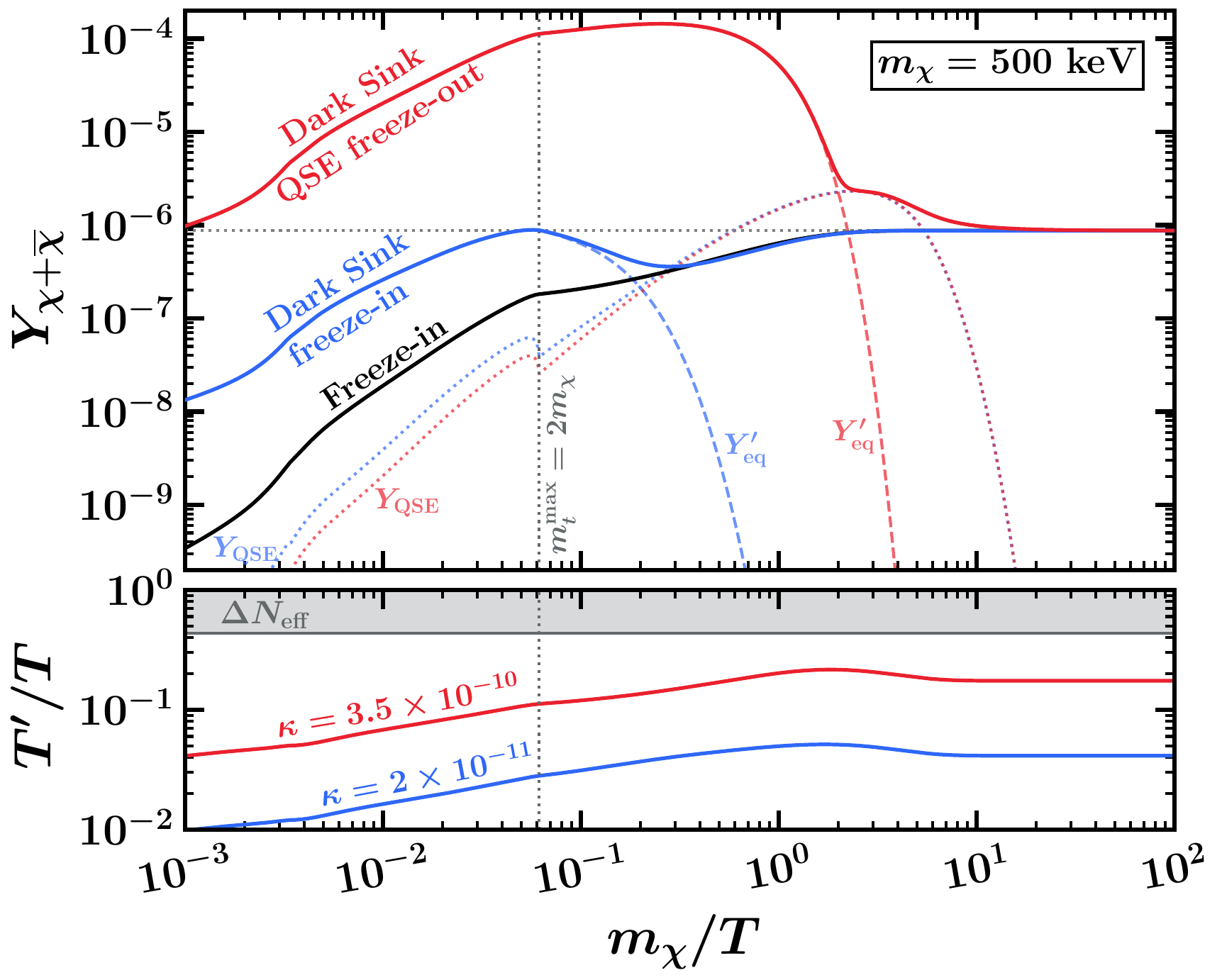}
  \end{minipage}
  \begin{minipage}[]{\columnwidth}
    \centering
    \includegraphics[width=0.8\columnwidth]{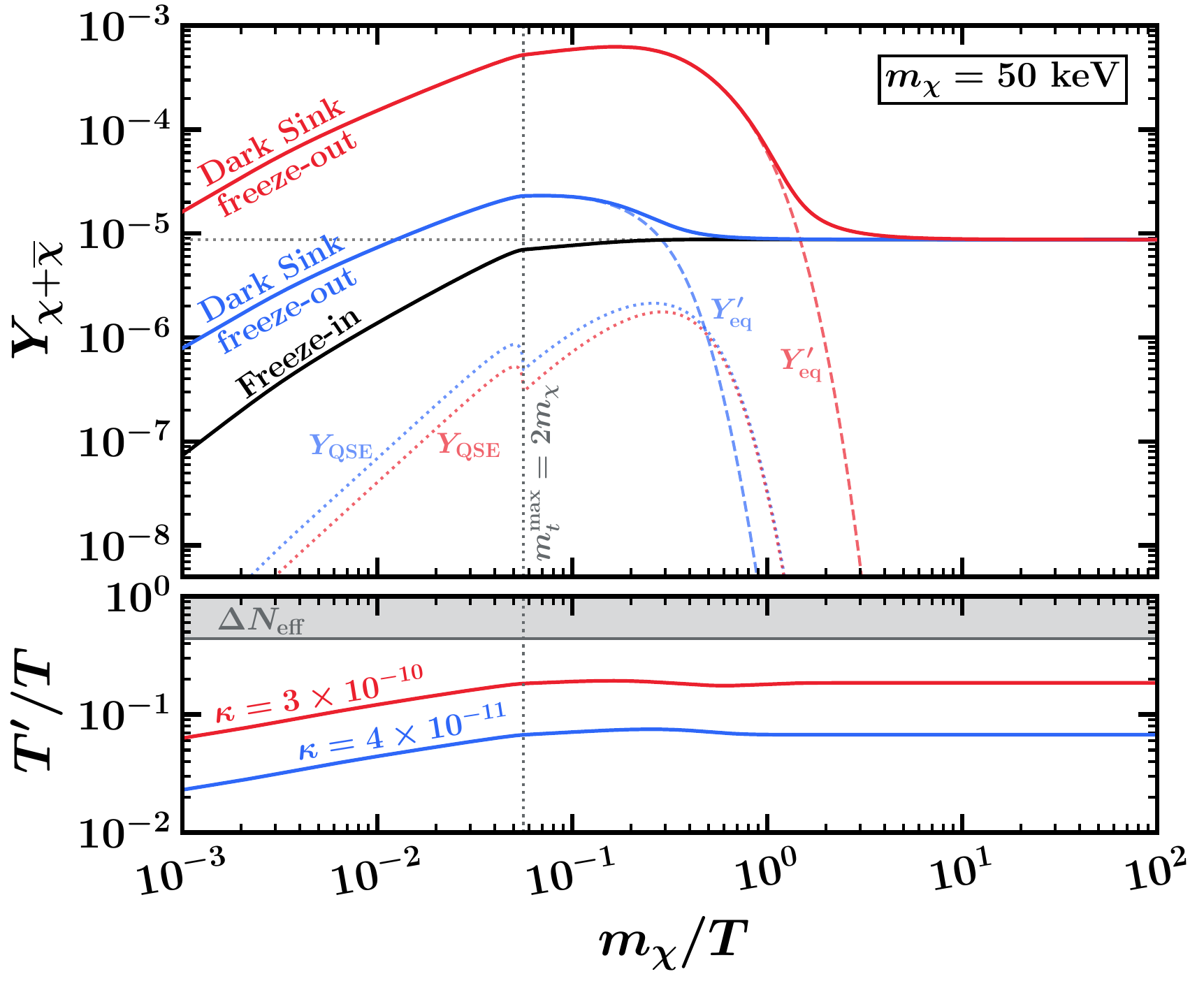}
  \end{minipage}
 \caption{Evolution of dark matter yield $Y_{\chi+ \overline{\chi}}$ and the ratio of the dark-sector and the SM-bath temperatures $T'/T$ for $m_\chi = 500$ keV (top) and $m_\chi = 50$ keV (bottom) 
  as a function of $m_\chi/T$. Freeze-in values for $\kappa$ (without the Dark Sink) are $1.76 \times 10^{-11}$ and $\kappa_{FI}= 3.47 \times 10^{-11}$ for 500 keV and 50 keV, respectively.   
\label{fig:Yield}}
\end{figure*}

In Fig.~\ref{fig:Yield}, we show the evolution of the dark matter yield $Y_{\chi+\overline{\chi}}$ and the ratio of the dark-sector temperature to the visible-sector temperature $T^{\prime}/T$ as a function of $m_{\chi}/T$ for two different dark matter masses (500 keV, 50 keV).  These masses are chosen because they capture the essential features of the Dark Sink scenario in different areas of the currently viable parameter space. The horizontal dotted gray lines in the yield plots correspond to the dark matter relic abundance observed today. 
\beq
m_\chi Y_{\chi+\overline{\chi}} = \frac{\Omega_{\chi+\overline{\chi}} \rho_\text{crit}}{s_0}
= 4.37 \times 10^{-10} \text{ GeV}.
\label{eq:relic}
\eeq
Here, $\Omega_{\chi+\overline{\chi}}$ is the dark matter energy fraction, $\rho_\text{crit}$ is the critical energy density, $s_0$ is the entropy density of the photon bath today, and $Y_{\chi+\overline{\chi}}$ is the dark matter yield.
The vertical gray dotted lines denote the temperatures below which plasmons are kinematically forbidden from decaying to dark matter. These are the temperatures at which the maximum values of the transverse plasmon masses $m_t^\text{max}$ equal $2 m_\chi$ (see Fig.~\ref{fig:PlasmonFITemperature}).

Before describing the behavior at each of these benchmark masses in more detail, we first discuss general features shared by all Dark Sink scenarios.  At early times the dark matter yield tracks the equilibrium yield for a particle in contact with a thermal bath at dark-sector temperature $T^{\prime}$.  To understand this, note that at early times the quasi-static equilibrium yield defined in Eq.~(\ref{eq:yieldQSE}) is much smaller than the equilibrium yield $Y_{\mathrm{QSE}}\ll Y_{\mathrm{eq}}^{\prime}$.  When this is true, the quasi-static equilibrium term can be dropped from the Boltzmann Eq.~(\ref{eq:FullBoltSimp}), and $Y$ is driven to $Y_{\mathrm{eq}}^{\prime}$.  The precise manner in which the dark matter yield eventually departs from $Y_{\mathrm{eq}}^{\prime}$ depends on which terms in Eq.~(\ref{eq:FullBoltSimp}) come to dominate. This differs for different values of $\kappa$ and $m_{\chi}$; we will discuss examples below. 

But first, we note that at early times, for all $m_{\chi}$ and $\kappa$, the ratio $T^{\prime}/T$ is increasing as the temperature of the SM bath drops. This occurs for two reasons. First, recall that $T' < T$ (see Eq.~\eqref{eq:Tpbound}) is a strict inequality so net energy transfer is always from the SM bath to the dark sector. Second, at early times, $2\to2$ annihilations and plasmon decays are rapid relative to Hubble. So, energy is efficiently transferred to the dark sector; this in turn explains why $Y_{\mathrm{eq}}^{\prime}$ is increasing on the left side of Fig.~\ref{fig:Yield}. Eventually, for $T \sim m_{e}$, the rate of energy transfer loses out to Hubble expansion. Then both $T$ and $T^{\prime}$ decrease together as expected as the scale factor of the universe increases. Thus, $T^{\prime}/T$ levels out at late times. Examination of the $T'/T$ plots in Fig.~\ref{fig:Yield} reveals another feature at $T \sim m_{e}$, where the SM bath is heated relative to the dark sector as the electrons leave chemical equilibrium. 

For a fixed dark matter mass, increasing $\kappa$ increases the energy flow from the visible to the dark sector.  This means that increasing $\kappa$ increases $T^{\prime}/T$ and results in a more pronounced peak for $Y_{\mathrm{eq}}^{\prime}$.  Note, for a fixed $m_{\chi}$, $\rho^\prime/\rho$ scales as $\kappa^2$ (for $\rho^{\prime} \ll \rho)$, see Eqs.~(\ref{eq:rhoprimeBoltzmann})-(\ref{eq:collisionPlasmon}). Consequently, asymptotic values of $T^{\prime}/T$ scale as $\sqrt{\kappa}$ for fixed $m_{\chi}$. 
The bounds on $\Neff$ from Planck's measurements of the CMB and measurements of baryon acoustic oscillations (BAO) \cite{Planck:2018vyg,Cielo:2023bqp} require $\Delta \Neff < 0.28$ at 95 \% CL. At the time of last scattering, only relativistic $\psi$'s contribute to $\Neff$ and the final dark-sector temperature is thus bounded above by
\begin{equation}
\label{eq:Tpbound}
    T^{\prime}/T < 0.437\qquad \mathrm{(95\%\ CL)}.
\end{equation}
The $\Delta \Neff$ constraint from Eq.~\eqref{eq:Tpbound} is shown in gray, and sets an upper limit on $\kappa$ for a given $m_{\chi}$, though as we will see in Sec.~\ref{sec:xsec}, bounds from direct detection are stronger.

Now, we turn to the details of the dark matter yield evolution.  We first focus on the $m_{\chi} = 500$ keV case, shown in the top half of Fig.~\ref{fig:Yield}.  The traditional freeze-in benchmark (with no Dark Sink) that accounts for plasmon decays and $2 \rightarrow 2$ processes is shown in black.  The corresponding portal coupling is $\kappa = 1.76 \times 10^{-11}$.  The yield gradually increases as the temperature decreases until the temperature drops to a value that is too low to produce additional dark matter. The blue and red curves are Dark Sink benchmarks. The red curve, labeled ``Dark Sink QSE freeze-out" corresponds to a substantially larger $\kappa$ than the freeze-in benchmark.  In this case, the dark matter yield at early times exceeds its asymptotic value. The dominant contribution to the right-hand side of the Boltzmann Eq.~(\ref{eq:FullBoltSimp}) is provided by  $Y^{\prime}_\text{eq}$ and so $Y$ traces the equilibrium abundance in the dark sector (dashed red curve labeled $Y^{\prime}_\text{eq}$).  At early times, since $T^{\prime}/T$ is increasing, as discussed above, this dark-sector equilibrium value increases even as the visible temperature decreases. However, once the dark matter mass exceeds the dark-sector temperature, $Y^{\prime}_\text{eq}$ begins to drop exponentially. Eventually, the quasi-static equilibrium yield given by Eq.~(\ref{eq:yieldQSE}) becomes comparable-to and then larger-than the dark-sector equilibrium yield $Y^{\prime}_\text{eq}$.  During this time, the $Y_\text{QSE}$ term provides the dominant contribution on the right-hand side of Eq.~(\ref{eq:FullBoltSimp}) and the dark matter yield is driven to $Y_\text{QSE}$.  Finally, all reaction rates are swamped by the Hubble expansion rate and the yield settles to its observed value. This occurs when 
\begin{equation}
\label{eq:QSFOcond}
n^\text{eq}_\chi \sqrt{\langle \sigma v \rangle_\text{eff} \langle \sigma v \rangle_{\chi \overline{\chi} \to \psi \overline{\psi}}^{\prime}} \sim H, \qquad  \text{(Dark Sink QSE freeze-out)}
\end{equation}
where $n^\text{eq}_\chi$ is the dark matter equilibrium number density at SM bath temperature $T$.

On the other hand, the blue curve labelled ``Dark Sink Freeze-in" corresponds to a $\kappa$ value $2 \times 10^{-11}$ that is only slightly above the freeze-in value of $\kappa_{FI} = 1.76 \times 10^{-11}$. As is the case for all Dark Sink scenarios, at early times the yield tracks $Y^{\prime}_\text{eq}$. However, the smaller value of $\kappa$ in this case means that $T^{\prime}/T$ is smaller, and the peak of the $Y^{\prime}_\text{eq}$ (and, correspondingly, the amount of dark matter present) is still smaller than the required yield at late times. Because of this, once $Y^{\prime}_\text{eq}$ begins to drop exponentially, the first two terms on the right-hand side of Eq.~(\ref{eq:FullBoltSimp}) are small and the whole right-hand side of the Boltzman equation is well approximated by $(n_\chi^{\rm eq})^2\langle \sigma v \rangle_{\mathrm{eff}}$ -- exactly what would be expected in the typical freeze-in scenario. Importantly, in the transition from the $Y^{\prime}_\text{eq}$ curve to what looks like a typical freeze-in curve, the dark matter yield initially slightly undershoots the traditional freeze-in yield, thus achieving the observed dark matter abundance ultimately requires a slightly larger value of $\kappa$.

In the lower half of Fig.~\ref{fig:Yield}, we show results for $m_{\chi} = 50 \text{ keV}$. In this case, even for values of $\kappa$ just above the freeze-in value (the blue curve, $\kappa= 4 \times 10^{-11}$, compare the freeze-in value of $\kappa = 3.47 \times 10^{-11}$), the dark-sector equilibrium curve $Y^{\prime}_\text{eq}$ already has a peak that exceeds the asymptotic value.  Furthermore, the quasi-static equilibrium yield (dashed curve, $Y_\text{QSE}$) is never the dominant contribution to the Boltzmann equation. So, for this mass value, the dark matter yield just follows the dark-sector equilibrium yield until it undergoes freeze-out
when 
\begin{equation}
\label{eq:DSFOcond}
n_\chi^{\mathrm{eq} \,\prime} \langle \sigma v \rangle_{\chi \overline{\chi} \to \psi \overline{\psi}}^{\prime} \sim H, \qquad \text{(Dark Sink freeze-out)}
\end{equation}
where $n_\chi^{\mathrm{eq} \,\prime}$ is the dark matter equilibrium number density at dark-sector temperature $T'$. This behavior persists to larger values of $\kappa$; an example with $\kappa= 3 \times 10^{-10}$ is shown as the red curve. 

\begin{figure*}[t!]
  \begin{minipage}[]{\columnwidth}
    \centering
    \includegraphics[width=0.8\columnwidth]{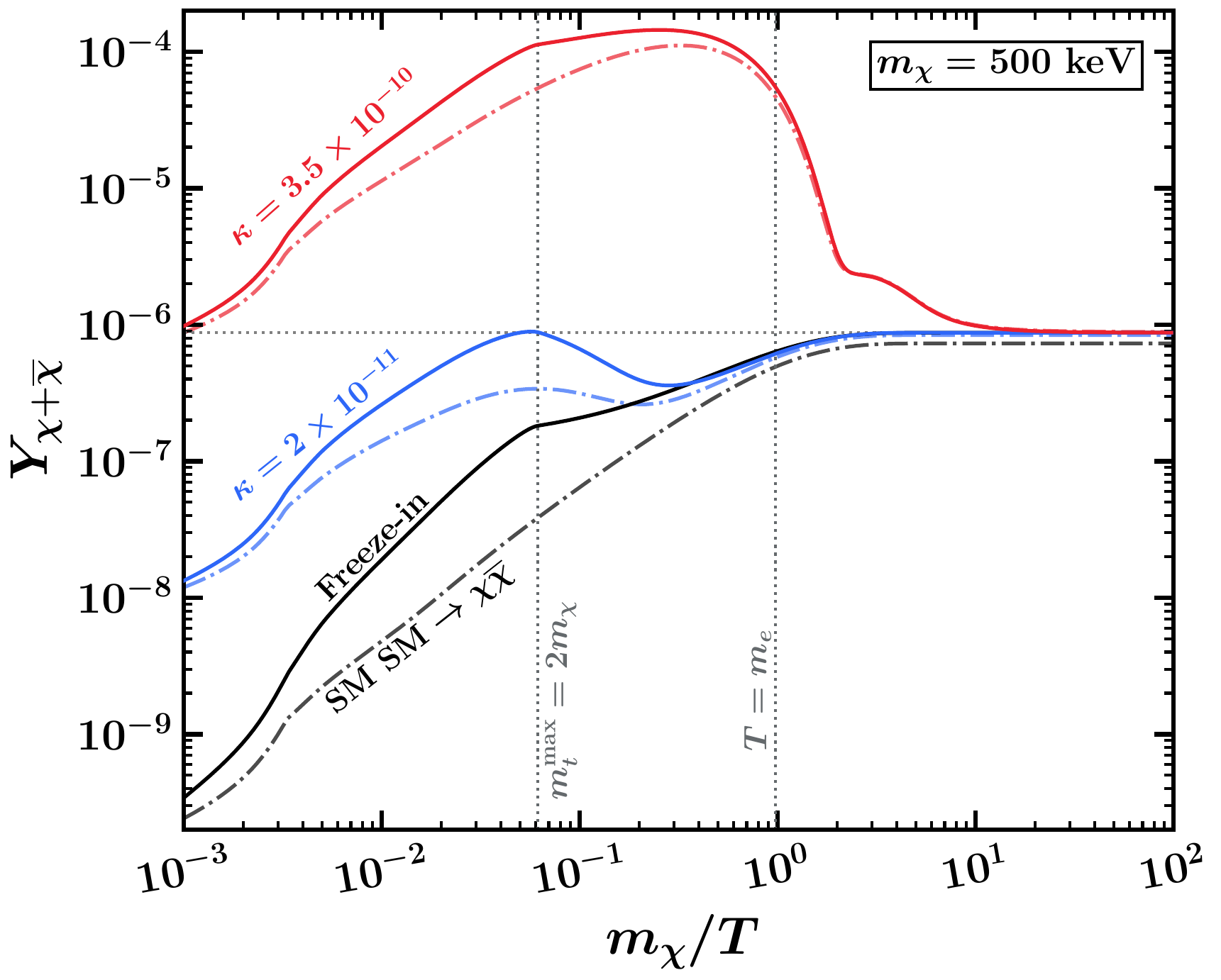}
  \end{minipage}
  \begin{minipage}[]{\columnwidth}
    \centering
    \includegraphics[width=0.8\columnwidth]{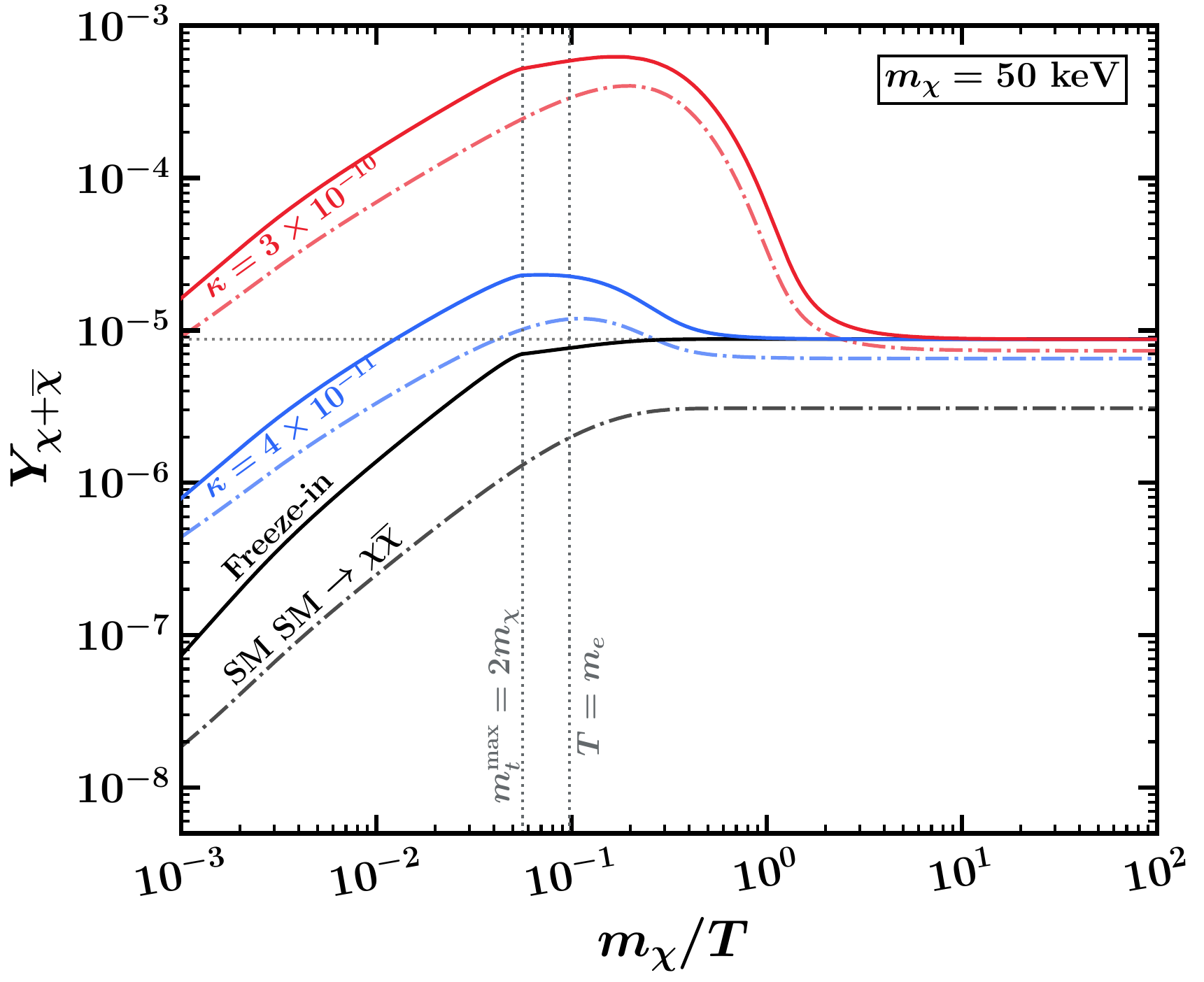}
  \end{minipage}
 \caption{Evolution of dark matter yield in the dark sector as a function of $m_{\chi}/T$ for various values of $\kappa$, as labelled, for $m_{\chi} = 500$ keV (top) and 50~keV (bottom). The yield using only the 2 $\rightarrow$ 2 process is shown as a dot-dashed line, while the full yield is shown as a solid curve.  
\label{fig:PlasmonEffect}}
\end{figure*}

In Fig.~\ref{fig:PlasmonEffect}, we show the evolution of the yield as a function of $m_{\chi}/T$ for the same benchmarks of $m_{\chi}$, 500 keV (upper panel) and 50 keV (lower panel), and $\kappa$. In this case, the solid curves are the full result, while the dot-dashed curves correspond to the evolution that would occur in the presence of the $2 \rightarrow 2$ process alone (i.e. without the plasmon contribution). The two vertical gray dotted lines correspond to temperatures below which dark matter freeze-in from plasmon decays (the $T$ where $m_t^\text{max} = 2 m_\chi$; see Fig.~\ref{fig:PlasmonFITemperature}) and the $2 \rightarrow 2$ process ($T \sim m_e$) are kinematically forbidden and Boltzmann suppressed, respectively. The larger gap between these two sets of curves in the case of 50 keV highlights the relative importance of plasmon decays at lighter masses. Also notable is that the effects of plasmon decay are most relevant for lower values of $\kappa$ closer to the freeze-in benchmark (i.e. where the effects of the Dark Sink are least pronounced). In the absence of the Dark Sink, the dark matter abundance is entirely controlled by interactions with the SM bath. When the Dark Sink is turned on, the dark-sector interactions also play an important role in establishing the dark matter yield. Since the addition of plasmons impacts the interactions between the SM and the dark matter, a bigger effect is observed in the absence of the Dark Sink. 

\subsection{Dark Sink Interactions}
\label{sec:GPhi}

In this section, we discuss the Dark Sink annihilation rates necessary to achieve the observed dark matter abundance. For fixed dark matter mass $m_\chi$ and portal coupling $\kappa$, imposing that the right yield is achieved, Eq.~\eqref{eq:relic}, determines the requisite Dark Sink Fermi constant $G_\Phi$. 

\begin{figure*}[t!]
  \begin{minipage}[]{\columnwidth}
    \centering
    \includegraphics[width=0.75\columnwidth]{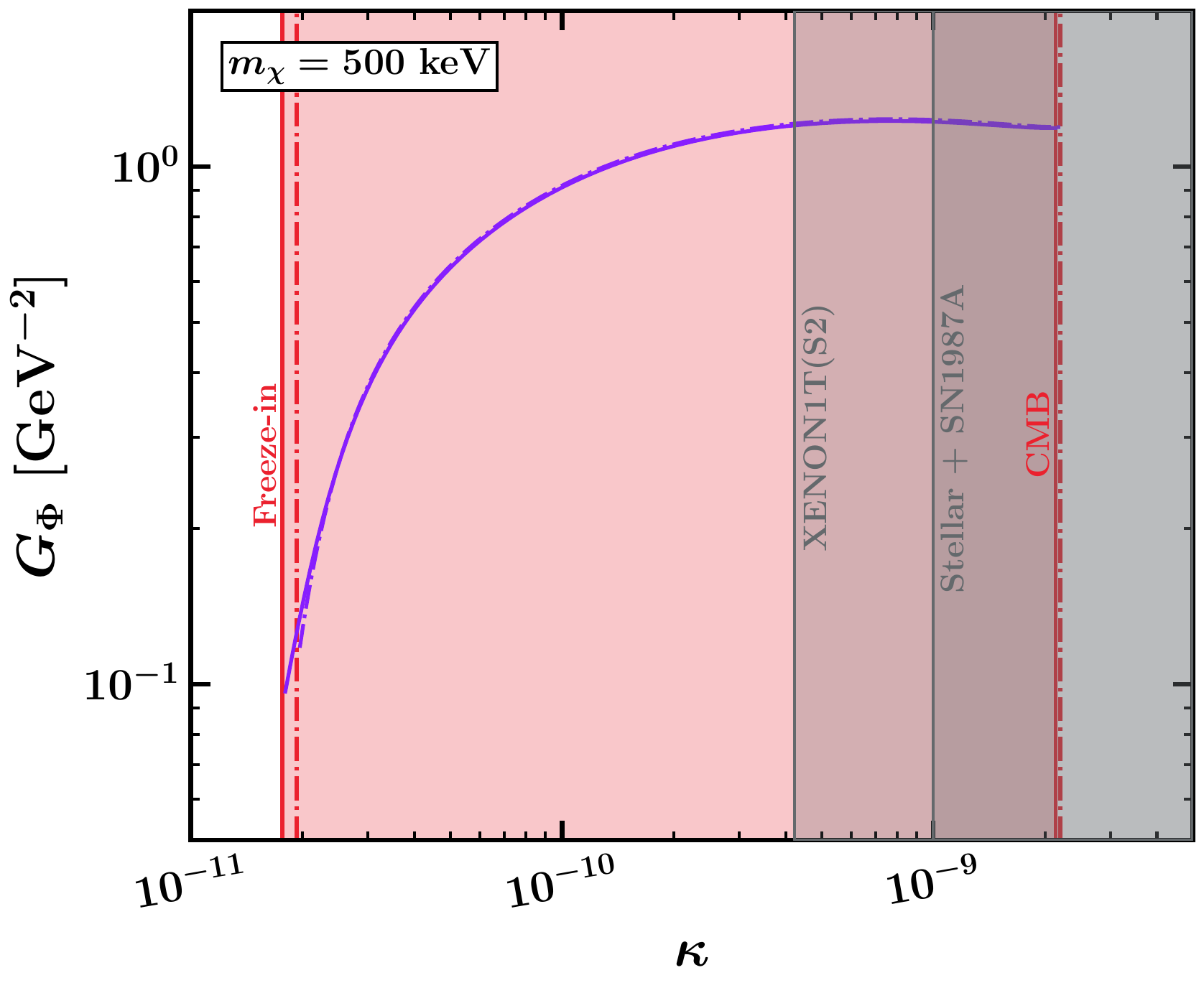}
  \end{minipage}
  \begin{minipage}[]{\columnwidth}
    \centering
    \includegraphics[width=0.75\columnwidth]{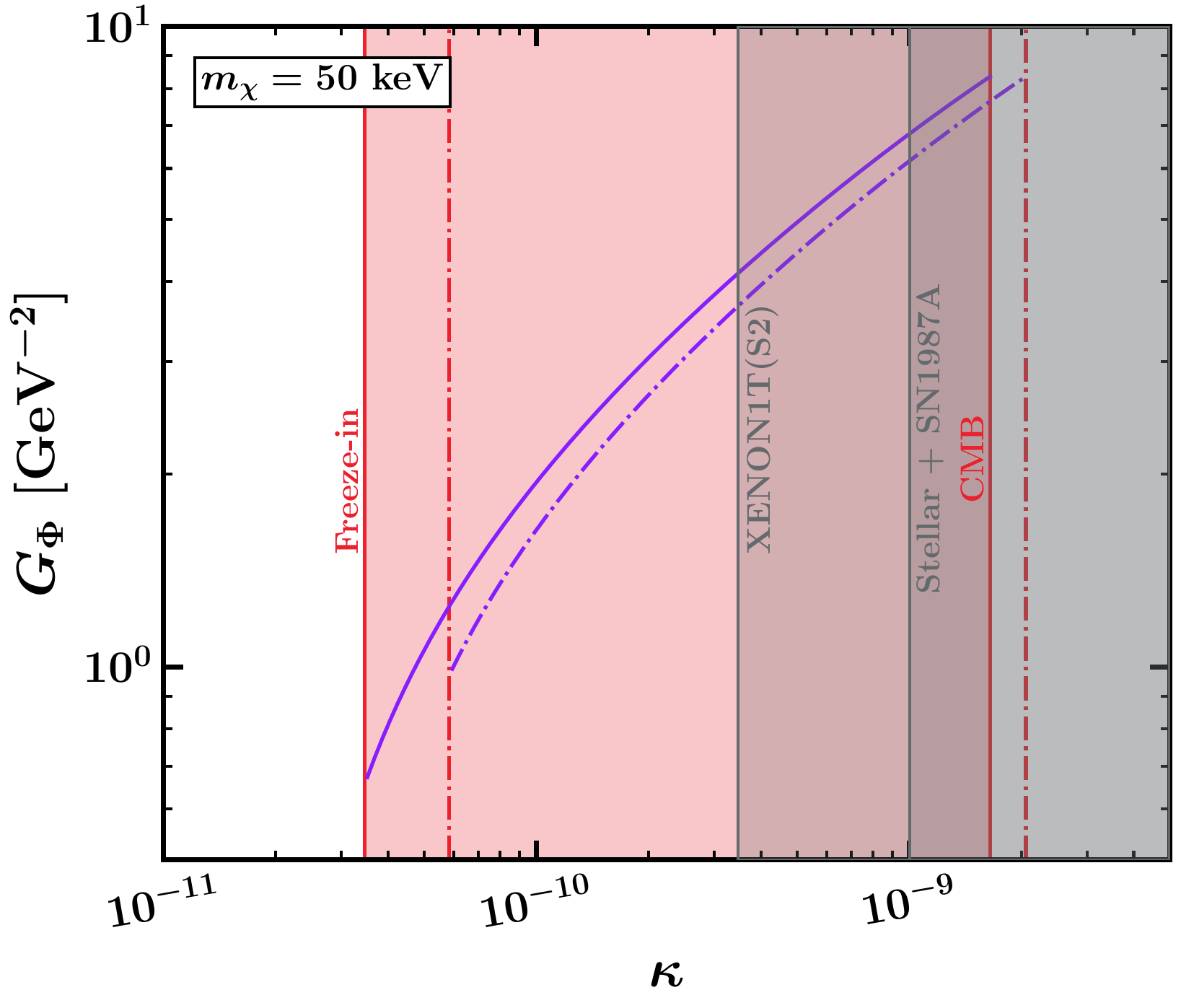}
  \end{minipage}
 \caption{The Dark Sink Fermi constant $G_\Phi$ that reproduces the required dark matter abundance as a function of $\kappa$ for dark matter masses of 500 keV (top panel) and 50 keV (bottom panel) with (solid purple lines) and without (dot-dashed purple lines) plasmons. See text for details. 
\label{fig:GPhi}}
\end{figure*}

In Fig.~\ref{fig:GPhi}, we show the $G_{\Phi}$ that realizes the correct dark matter abundance as a function of $\kappa$ for dark matter benchmark masses of 500 keV (top) and 50 keV (bottom). Solid curves show the $G_{\Phi}$ in our full calculation, while the dot-dashed curves give the values of $G_{\Phi}$ when we neglect plasmons (i.e., only accounting for $\text{SM SM} \rightarrow \chi \overline{\chi}$ processes). Notably, the two curves are difficult to distinguish for $m_{\chi} = 500$ keV.  This is indicative of the small role plasmons play for $m_\chi \gtrsim \text{MeV}$. The difference is more pronounced for 50 keV.  The shaded red regions in the plots book-ended by the curves marked ``Freeze-in" and ``CMB" indicate values of $\kappa$ that are achievable in the Dark Sink scenario.  The lowest value of $\kappa$ just reproduces the case without a Dark Sink.  The highest value corresponds to the case where so much energy is moved from the visible to the dark sector that subsequent annihilations of dark matter to $\psi$'s produce a too-large amount of dark radiation, thus violating the $N_\text{eff}$ constraint.  

The solid red boundaries are those that result from the full calculation, while the dot-dashed boundaries would result if plasmons were neglected. Shaded gray regions violate current direct detection \cite{XENON:2019gfn,An:2021qdl} (marked XENON (S2)) or astrophysical \cite{Vogel:2013raa,Chang:2018rso} bounds from stellar cooling or SN1987A \cite{Vogel:2013raa,Chang:2018rso}. 

In the cases where the dark matter abundance is determined by freeze-out from $Y_\text{QSE}$ or $Y^\prime_\text{eq}$, it follows from the Boltzmann Eq.~(\ref{eq:FullBoltSimp}) that the dark matter yield at freeze-out is $Y_{\chi+\overline{\chi}} \sim 2 \, H/(s \langle \sigma v \rangle_{\chi \overline{\chi} \to \psi \overline{\psi}}^{\prime})$. This, together with Eqs.~(\ref{eq:relic}) and (\ref{eq:sigmavprimelargex}), implies 
\begin{equation}
G_\Phi
\approx
3.9  \text{ GeV$^{-2}$}
\,
\left(
\frac{100 \text{ keV}}{m_\chi}
\right)
\left(
\frac{\sqrt{g_{\ast}}}{g_{\ast,s}}
\right)^{1/2}
\left(
\frac{\sqrt{x_f x_f^\prime}}{5}
\right)
,
\label{eq:Gphiscaling}
\end{equation}
where $g_{\ast}$ is the relativistic degrees of freedom and $x_f^{(\prime)} \equiv m_\chi/T_f^{(\prime)}$, with the subscript $f$ corresponding to quantities at freeze-out (from $Y_\text{QSE}$ or $Y^\prime_\text{eq}$).

The scaling of $G_\Phi$ as a function of $\kappa$ shown in Fig.~\ref{fig:GPhi} for a fixed $m_\chi$, can be understood using Eq.~(\ref{eq:Gphiscaling}) as follows. The ratio $T^\prime_f/T_f$, or equivalently $x_f/x_f^\prime$, scales like $\sqrt{\kappa}$ as shown in Fig.~\ref{fig:Yield} and discussed in Sec.~\ref{sec:evolution}. The final ingredient in determining the scaling of $G_{\Phi}$ with respect to $\kappa$ is an understanding of the scaling of $x_{f}$ (or $x_f^{\prime}$). In the Dark Sink freeze-out regime in which dark matter freezes-out from $Y^\prime_\text{eq}$, the scaling of $x_f$ (or $x_f^\prime$) versus $\kappa$ can be obtained from Eq.~\eqref{eq:DSFOcond}. This is the case for $m_\chi = 50$ keV (and $m_\chi = 500$ keV for $\kappa \gtrsim 2 \times 10^{-9}$). In contrast, in the Dark Sink QSE freeze-out regime in which dark matter freezes-out from $Y_\text{QSE}$, the $x_f^{(\prime)}$ scaling can be obtained from Eq.~\eqref{eq:QSFOcond}. This is the case for $m_\chi = 500$ keV for $3 \times 10^{-11} \lesssim \kappa \lesssim 2 \times 10^{-9}$. It is also interesting that for this range of $\kappa$, the requisite $G_{\Phi}$ with the inclusion of plasmons are slightly \emph{smaller} than without as seen in the top of Fig.~\ref{fig:GPhi}. At first, this might be surprising. After all, the naive expectation might be that since plasmon decays contribute an additional source of dark matter, a correspondingly larger $G_{\Phi}$ would be needed to annihilate it away. However, in the case where the dark matter abundance is determined by freeze-out from quasi-static equilibrium, the dominant effect of the plasmons is not this extra dark matter production, but rather an increase in the temperature of the dark bath $T^{\prime}$. This is evident from Eq.~(\ref{eq:Gphiscaling}) where $T_f^\prime$ with plasmons would be slightly larger than without, leading to this effect.

\subsection{Direct Detection} 
\label{sec:xsec}

\begin{figure}[t!]
\centering
\includegraphics[width=0.98 \columnwidth]{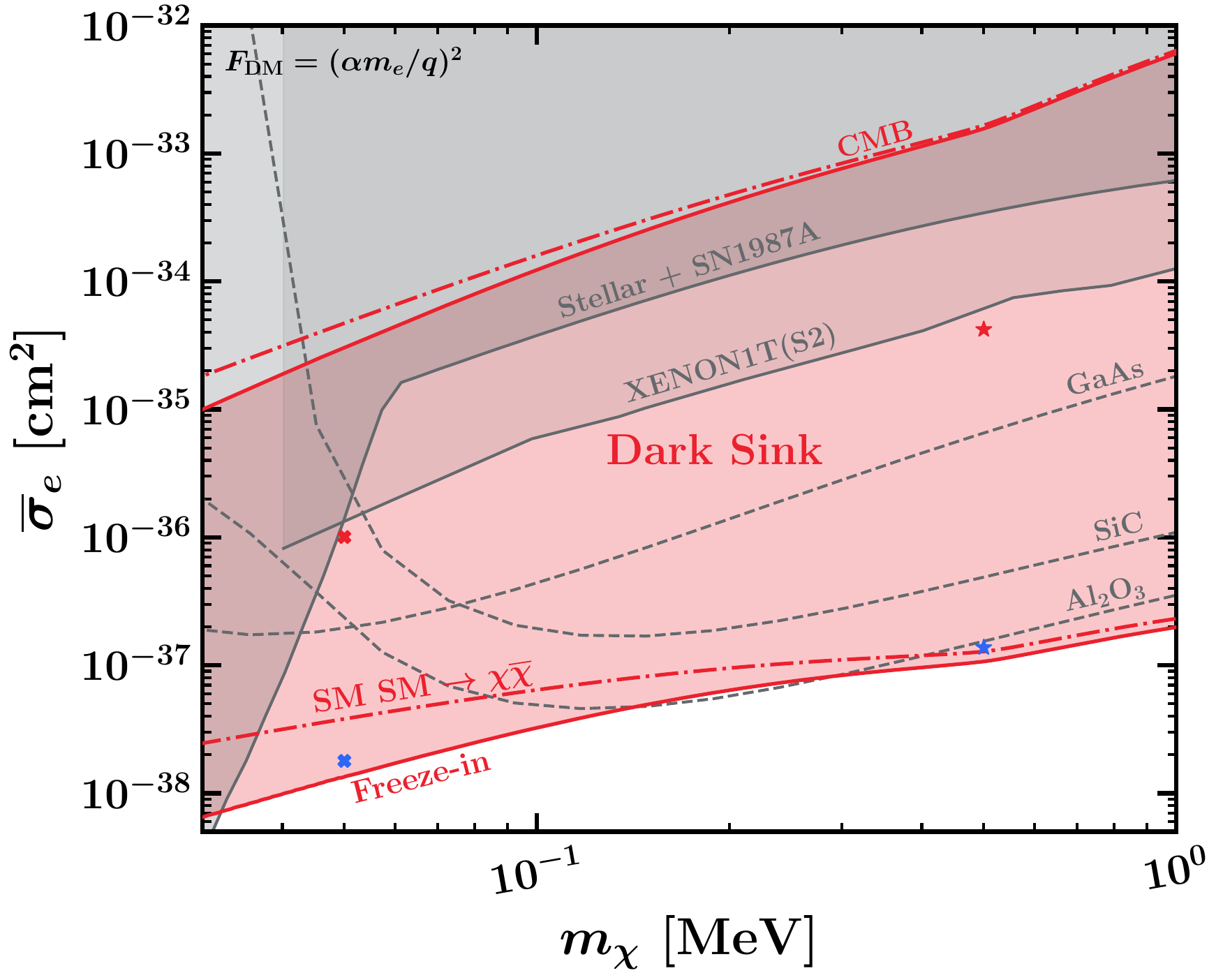}
\caption{The introduction of the Dark Sink allows the entire {\color[HTML]{EB212E} red} shaded region to reproduce the correct relic abundance.  The bottom of the red region corresponds to the usual freeze-in benchmark in the absence of the Dark Sink.  The top of the red region, marked `CMB' corresponds to cross sections where the amount of dark radiation exceeds current bounds at the 95\% CL \cite{Planck:2018vyg,Cielo:2023bqp}.  The dot-dashed curves correspond to an analysis neglecting the effect of plasmons. The points marked as $\times$ ($\star$) correspond to $m_\chi=50 \text{ keV}$ ($m_\chi=500 \text{ keV}$) benchmarks whose evolution histories are shown in Fig.~\ref{fig:Yield}. Current direct detection \cite{XENON:2019gfn,An:2021qdl} and astrophysical \cite{Vogel:2013raa,Chang:2018rso} bounds are shown in solid gray, while the projected bounds at the 95\% CL with 1 g-day exposure and zero background are shown in dotted gray \cite{Knapen:2021bwg}. See text for more details.
}
\label{fig:SubMeVDarkSink}
\end{figure}

In the absence of the Dark Sink, a single value of $\kappa$ yields the observed relic abundance for a given dark matter mass. Since $\kappa$ also controls the scattering of dark matter off electrons, this freeze-in value of $\kappa$ predicts a corresponding direct detection cross section: \cite{Essig:2015cda}
\begin{eqnarray} \label{eq:DDXsection}
\overline{\sigma}_e &=& \frac{16 \pi \mu_{\chi e}^2 \alpha^2 \kappa^2}{(\alpha m_e)^4}\\
&=& 5.4 \times 10^{-39} \; \rm{cm}^2 \; \left(\frac{\kappa}{10^{-10}} \right)^2 \left(\frac{\mu_{\chi e}}{10^{-2} \; \rm{MeV}} \right)^2 \nonumber ,
\end{eqnarray}
where we have evaluated the cross section at the momentum transfer $q= \alpha m_{e}$.  Plotting the value of the direct detection cross section that reproduces the observed dark matter abundance as a function of $m_{\chi}$ gives the typical freeze-in curve \cite{Dvorkin:2019zdi}, reproduced in Fig.~\ref{fig:SubMeVDarkSink}. The existing code repository {\sc FreezeIn} \cite{FreezeInCode} has been updated to account for plasmon decays in the predictions for the freeze-in model.

In the presence of the Dark Sink, the extra dark matter annihilation channel parameterized by $G_{\Phi}$ instead allows a range of $\kappa$ values to yield the correct relic abundance for a given dark matter mass. This in turn corresponds to a range of viable direct detection cross sections, shown in Fig.~\ref{fig:SubMeVDarkSink} as a function of the dark matter mass. The red-shaded region corresponds to cross sections compatible with the Dark Sink model presented here. The lower boundary corresponds to the traditional freeze-in benchmark in the absence of the Dark Sink.  And while higher cross sections (larger $\kappa$) would lead to an over-production of dark matter in the absence of the Dark Sink, the Dark Sink allows the extra energy density in the dark sector to be converted to dark radiation $\psi$. As discussed in Eq.~(\ref{eq:Tpbound}), the amount of energy density in dark radiation is bounded by cosmological measurements.  The upper boundary of the red-shaded region in Fig.~\ref{fig:SubMeVDarkSink} reflects this constraint. 

The dot-dashed red lines near the boundaries of the red-shaded region are the would-be boundaries of the Dark Sink scenario if one were to neglect the effects of the plasmon decays. That is, the lower dot-dashed red line is the freeze-in benchmark, while the upper line would correspond to the Dark Sink scenario that saturates the $\Neff$ bound, both in the absence of plasmon decays. At masses near an MeV, the difference between the plasmon and no-plasmon curves is small, while at lower masses, the effects are more pronounced. Moreover, the lower (i.e. freeze-in benchmark) and upper boundary (i.e. $\Neff$ bound) are not impacted equally. This is because the shift in the lower boundary depends upon how the plasmons impact the dark matter \emph{number} density, whereas the shift in the upper boundary depends on how plasmons impact the flow of \emph{energy} to the dark sector.  Additionally, the upper boundary is a case where the dark matter is typically produced via a Dark Sink freeze-out process, whereas the lower boundary corresponds to a freeze-in process.

The solid gray lines/regions in Fig.~\ref{fig:SubMeVDarkSink} show  current experimental bounds from a variety of astrophysical \cite{Vogel:2013raa,Chang:2018rso} and direct detection \cite{XENON:2019gfn,An:2021qdl} experiments. Below $\sim 50$ keV the stellar \cite{Vogel:2013raa} and supernova \cite{Chang:2018rso} constraints are most stringent, while above  $\sim 50$ keV the leading constraints are provided by XENON1T S2 data \cite{XENON:2019gfn} from solar reflected dark matter \cite{An:2021qdl}. The gray dashed lines in Fig.~\ref{fig:SubMeVDarkSink} show the expected reaches of a variety of next-generation direct detection experiments. As a representative of future experiments, we show the 95\% CL expected reaches for various polar materials (GaAs, SiC, and Al$_2$O$_3$ \cite{Knapen:2017ekk,Knapen:2021bwg}) with 1 g-day exposure and zero background.\footnote{The expected reaches in the given references are for a 1 kg-year exposure. To obtain the reaches for a g-day exposure, we scale the curves by a factor of $365 \times 10^3$, appropriate for zero background.} Additionally, there are other experiments that rely on doped semiconductors (Si:P \cite{Du:2022dxf}), superconductors (Al \cite{Hochberg:2015pha,Hochberg:2015fth}), and Dirac materials (ZrTe$_5$ \cite{Hochberg:2017wce}). Excitingly, XENON1T is currently constraining the Dark Sink scenario in this mass range, and further improvements in the bounds, no matter how modest, will continue to probe Dark Sink models. In the mid-term, as new experiments come online with very small initial exposures, they will be able to meaningfully test the sub-MeV Dark Sink on the road to the freeze-in line.

\section{Conclusions}
We have presented a concrete model for dark matter with mass as light as 10's of keV with a direct detection cross section orders of magnitude larger than the freeze-in benchmark. Its key distinction relative to the standard freeze-in scenario is the presence of a Dark Sink that shuttles energy away from dark matter and into dark radiation. Modest improvements in current direct detection experiments will probe this model further, while new promising experiments will probe it fully.

We have emphasized the importance of plasmon decays in the sub-MeV regime. The plasmon decays become increasingly important at lower masses, and dominate the production for the lowest masses of interest, ${\mathcal O}(10$ keV). Plasmons directly impact the strength of Dark Sink annihilations required to reproduce the observed relic density.  

While we have focused on a particular Dark Sink interaction mediated by a heavy scalar $\Phi$, other Dark Sink models exist, and the calculations here provide a framework for their analyses. It would be of interest to investigate the impact of a Dark Sink in other freeze-in benchmarks.
 
\begin{acknowledgments}
This material is based upon work in part supported (A.P.) by the U.S. Department of Energy, Office of Science, Office of High Energy Physics under Award Number grant DE-SC0007859. This research was supported in part through computational resources and services provided by Advanced Research Computing (ARC), a division of Information and Technology Services (ITS) at the University of Michigan, Ann Arbor.
\end{acknowledgments}

\end{document}